\documentclass{article}
\usepackage{amsthm}
\usepackage{amsmath}
\usepackage{amssymb}
\usepackage{graphicx}
\usepackage{varwidth}
\usepackage{algpseudocode}
\usepackage{algorithm}
\usepackage{color}
\usepackage{tikz}
\usepackage{graphicx}
\usepackage{lscape}
\usepackage{rotating}
\usepackage{ulem}
\usepackage{subcaption}
\usetikzlibrary{arrows}

\theoremstyle{definition}

\theoremstyle{remark}

\begin{document}

\title{A Comparison of Approaches for Finding Minimum Identifying Codes on Graphs}
\author{Victoria Horan \thanks{\texttt{victoria.horan.1@us.af.mil}, Corresponding Author} \\  Air Force Research Laboratory \\ Information Directorate \\ \\ Steve Adachi \thanks{\texttt{steven.h.adachi@lmco.com}} \\ Lockheed Martin \\ \\ Stanley Bak \thanks{\texttt{stanley.bak.1@us.af.mil}} \\ Air Force Research Laboratory \\ Information Directorate}%


\date{\today}%

\maketitle

\let\thefootnote\relax\footnote{Approved for public release; distribution unlimited:  88ABW-2015-2163, DIS201511002.}

\begin{abstract}
In order to formulate mathematical conjectures likely to be true, a number of base cases must be determined.  However, many combinatorial problems are NP-hard and the computational complexity makes this research approach difficult using a standard brute force approach on a typical computer.  One sample problem explored is that of finding a minimum identifying code.  To work around the computational issues, a variety of methods are explored and consist of a parallel computing approach using Matlab, an adiabatic quantum optimization approach using a D-Wave quantum annealing processor, and lastly using satisfiability modulo theory (SMT) and corresponding SMT solvers.  Each of these methods requires the problem to be formulated in a unique manner.  In this paper, we address the challenges of computing solutions to this NP-hard problem with respect to each of these methods.
\end{abstract}

\section{Problem Statement and Background}

First introduced in 1998 \cite{first}, an {\it identifying code} for a graph $G$ is a subset of the vertices, $S\subseteq V(G)$,  such that for each $v\in V(G)$ the subset of vertices of $S$ that are adjacent to $v$ is non-empty and unique.  That is, each vertex of the graph is uniquely identifiable by the non-empty subset of vertices of $S$ to which it is adjacent.  More formally, let $N(v)$ be the set of vertices adjacent to $v$ and $B(v)=N(v) \cup \{v\}$.  Then we require that any two vertices $u,v \in V(G)$ have different identifying sets, or more precisely that we must have $S \cap B(v) \neq S \cap B(u)$, and also that both $S\cap B(v), S \cap B(u) \neq \emptyset$.  The combinatorial problem of finding minimum identifying codes has been shown to be NP-complete \cite{NPHard}, but also has many potential real-world applications.  The most commonly discussed application illustrates one use of identifying codes: placing sensors on a network.  If we place sensors on the nodes corresponding to our identifying code, then the set of sensors alerted gives us location information on the trigger.  For example, if we place smoke detectors in a house using an identifying code, then based off of which smoke detectors are set off we can pinpoint exactly which room of the house is on fire \cite{Sensors}.  Other applications appearing in the literature refer to similar scenarios, such as fault diagnosis of multiprocessor systems \cite{first}.

As some NP-complete problems are notorious for having special graph classes on which there are simple solutions, previous research has focused on the class of de Bruijn networks \cite{BoutinHoran}.  This paper explores the problem of finding the minimum size of an identifying code over the undirected de Bruijn graph using three different methods.  Section \ref{Dwave} describes an approach for solving the miminum identifying code problem using adiabatic quantum optimization \cite{Farhi}, which for small enough problem instances can be implemented on a D-Wave quantum annealing processor \cite{Johnson}.  While to our knowledge this is the first time adiabatic quantum optimization has been applied to the identifying code problem, it has been studied for other graph-theoretic problems including graph coloring \cite{Rieffel} and the graph isomorphism problem \cite{Gaitan}, \cite{Zick}; in addition, \cite{isingPaper} showed how to formulate a number of NP-complete problems as Ising models.  Other approaches are considered as follows:  Subsection \ref{Parallel} explores a parallel computing algorithm  and Subsection \ref{Stan} illustrates the method of using Satisfiability Modulo Theory (SMT) solvers.

While many of our examples and data revolve around the class of de Bruijn graphs, the methods discussed throughout can easily be applied to arbitrary graphs.  As this problem has not been considered before outside of \cite{UndirIDCodes}, no results exist on the minimum size of identifying codes on this class of graphs.  In this paper, we provide initial data on these values.  A summary of our complete contribution to these values is given in Figure \ref{SMTRes}.

For reference, we provide some of the basic definitions and background for the class of de Bruijn graphs here.  The undirected $d$-ary de Bruijn graph of order $n$, denoted $\mathcal{B}(d,n)$, is the graph with the following vertex and edge sets.

\begin{eqnarray*}
V & = & \{x_1x_2 \ldots x_n \mid x_i \in \{0,1,\ldots , d-1\}\} \\
E & = & \{(x,y) \mid x_2x_3 \ldots x_n = y_1 y_2 \ldots y_{n-1} \hbox{ or } x_1x_2\ldots x_{n-1} = y_2 y_3 \ldots y_n\}
\end{eqnarray*}

\begin{figure}
\begin{center}
\begin{tikzpicture}[-,>=stealth',auto,node distance=2cm,
  thick,main node/.style={circle,draw,font=\sffamily\bfseries,scale=0.75},new node/.style={circle,fill=black,text=white,draw,font=\sffamily\bfseries,scale=0.75}]

  \node[main node] (0) {000};
  \node[main node]  (1) [above right of=0] {001};
  \node[main node] (2) [below right of=1] {010};
  \node[main node]  (4) [below right of=0] {100};
  \node[main node] (5) [right of=2]       {101};
  \node[main node]  (6) [below right of=5] {110};
  \node[main node]  (3) [above right of=5] {011};
  \node[main node] (7) [below right of=3] {111};

  \path[every node/.style={font=\sffamily\small}]
    (0) edge node [left]      {} (1)
    (1) edge node [left]      {} (3)
        edge node [right]     {} (2)
    (2) edge node{} (5)
        edge node [right]     {} (4)
    (3) edge node [right]     {} (6)
        edge node [right]     {} (7)
    (4) edge node [left]      {} (0)
        edge node [right]     {} (1)
    (5) edge node [right]     {} (3)
    (6) edge node [right]     {} (5)
        edge node [right]     {} (4)
    (7) edge node [right]     {} (6);

\end{tikzpicture}
\end{center}
\caption{The 2-ary de Bruijn graph of order 3, or $\mathcal{B}(2,3)$.}\label{DBG}
\end{figure}
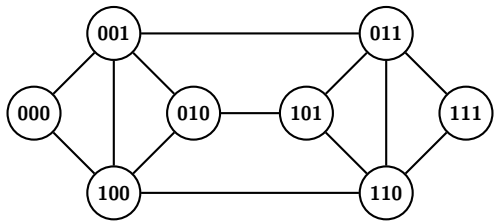

For example, the graph $\mathcal{B}(2,3)$ is illustrated in Figure \ref{DBG}.  These graphs have many useful properties for applications, such as having a relatively high number of nodes, a low degree at each node, and many short paths between any two nodes.  Additionally, many notoriously difficult problems such as the traveling salesman problem are solvable in polynomial time on this class of graphs.  For that reason, they have many interesting applications such as interconnection networks \cite{chips} and fault-tolerant wireless sensor networks \cite{WSN}.

\section{Quantum Annealing Approach}\label{Dwave}

  This section describes our approach for solving the minimum identifying code problem using quantum annealing.
Quantum annealing processors, such as those made by D-Wave Systems \cite{Johnson}, operate on the principle of adiabatic quantum optimization (AQO)\cite{Farhi}.
In AQO, the system Hamiltonian evolves according to the equation
\begin{equation} \label{eqn:adiabatic}
H(t) = \left(1-s\left(\frac{t}{T}\right)\right)H_{init} + s\left(\frac{t}{T}\right)H_{final},
\end{equation}
where $H_{init}$ is the initial Hamiltonian with a known and easily prepared ground state;  $H_{final}$ is the final Hamiltonian whose ground state corresponds to the solution of our optimization problem; and $s(\tau)$ increases from $s(0)=0$ to $s(1)=1$.
According to the adiabatic theorem \cite{Messiah}, for a large enough $T$ and a smooth enough function $s$ determined by the minimum spectral gap, a system starting in the ground state of $H_{init}$ at time $t=0$ will be in the ground state of $H_{final}$ at time $t=T$.
However, since a physical implementation of AQO does not strictly meet the conditions of the adiabatic theorem, in particular the assumption of a closed system,
we refer to the process as \textit{quantum annealing}, which may be viewed as a heuristic method for optimization.

The D-Wave processor is designed to solve quadratic binary minimization problems that can be expressed in terms of an \textit{Ising spin glass} energy functional.
Namely, a ground state of $H_{final}$ is a minimum of
\begin{equation} \label{eqn:ising}
\sum_j h_j S_j + \sum_{i \neq j} J_{ij} S_i S_j
\end{equation}
where $S_j = \pm1$ are the final qubit states in the computational basis, and the $h_j$ and $J_{ij}$ are programmable qubit biases and coupling strengths respectively.

While there are various ways of expressing the minimum identifying code problem as a binary minimization problem, we found that the most efficient approach, as measured by the number of qubits needed to solve the problem on the D-Wave processor, was to formulate the problem in terms of Boolean satisfiability.
This approach, described in Subsection \ref{subsec:SATForm}, leads to a conjunctive normal form (CNF) proposition containing clauses of various sizes.

Since the energy functional must be at most quadratic, we use ``gadgets'' (similar to those found in \cite{Gadgets}) to reduce the higher order clauses and generate an Ising model whose ground state encodes the solution to the satisfiability problem.
These gadgets add overhead in the form of ancillary binary variables that augment the problem variables; however we found that the overall qubit resource requirements for this approach are less than with other approaches we considered.
Our approach for mapping SAT clauses to an Ising model is described in Subsection \ref{subsec:SAT2Ising}.

The physical limitations of the D-Wave architecture present some additional challenges:
\begin{list}{$\bullet$}{}
	\item Not every pair of qubits on the chip are physically connected; rather, the connectivity can be represented as a square lattice of $K_{4,4}$ bipartite graphs, that D-Wave calls a \textit{Chimera graph}.
	Thus if the binary variables in the optimization problem are mapped 1-to-1 to qubits, not all of the quadratic coefficients $J_{ij}$ can be programmed.
	This challenge can be overcome using \textit{graph minor embedding} \cite{Choi2}, in which case a single binary variable may be mapped to multiple physical qubits.
	\item Furthermore, an actual D-Wave chip may contain a small number of faulty qubits (due to fabrication defects or calibration failures) which cannot be used in the graph minor embedding process \cite{Klymko}.
	While determining the optimal embedding for an arbitrary graph is itself an NP-hard problem, heuristic embedding techniques have been developed that provide reasonably efficient mappings \cite{Cai}.
	\item Finally, the D-Wave processor has a small amount of intrinsic control error (ICE), meaning that the actual values of biases and coupling strengths may differ slightly from the values programmed by the user.
	Some of the effects contributing to the ICE can be mitigated using \textit{gauge transformations} \cite{Boixo}, which are explained below.
\end{list}

Subsection \ref{subsec:Ising2DWave} describes the embedding and gauge transformation techniques we used to map the problem onto the D-Wave architecture.

The examples given in Subsections \ref{subsec:SATForm} through \ref{subsec:SolnUsingDWave} are for the undirected de Bruijn graph $\mathcal{B}(d,n)$ with $d=2$ and $n=4$.
This was the largest case we were able to solve on a D-Wave machine operated by Lockheed Martin and the University of Southern California, which had 504 working qubits.
In Subsection \ref{subsec:Scaling}, we also estimate the number of qubits that would be needed to solve the larger $d=2$ cases with $n=5,6$.
Compared to other types of optimization problems that have been studied using quantum annealing and the D-Wave machine, it appears that the qubit resource requirements for the minimum identifying code problem on de Bruijn graphs scale relatively well.

While the satisfiability-based approach proved to be the most efficient for this study, we also explored various approaches for expressing the minimum identifying code problem as a binary minimization problem.
For sake of completeness, Subsection \ref{subsec:OtherForms} describes two other approaches that we considered.

\subsection{Satisfiability Formulation}
\label{subsec:SATForm}

In this subsection, we describe how to formulate the minimum identifying code problem in terms of Boolean satisfiability.
While the method is presented here for a de Bruijn graph, the same approach can be applied to any graph.

We label the vertices of the de Bruijn graph $\mathcal{B}(d,n)$ in lexicographic order $i=0,1,2,...d^n-1$.
For a subset $S$ of $\mathcal{B}(d,n)$, we define Boolean variables
$$x_i = \left\{
\begin{array}{ll}
1, & \hbox{if $i\in s$;} \\
0, & \hbox{otherwise.}
\end{array}
\right.$$

We will construct a conjunctive normal form (CNF) formula that must be satisfied if $S$ is an identifying code.
Later, in the full optimization problem, we will add a penalty term of the form $\lambda \sum_i x_i$ to obtain the minimum identifying code.

For $S$ to be an identifying code, it must intersect the above-defined ball $B(i)$ for every $i$.
In other words, for every $i$ the clause
$$C_i = \bigvee_{j\in B(i)} x_j$$
must be satisfied for all $i$.
Similarly, the condition $S \cap B(i) \neq S \cap B(j)$ for all $i \neq j$ implies that
$$C_i \bigoplus C_j$$
must be satisfied for all $i \neq j$, where $\bigoplus$ is the exclusive OR operator.
Thus an identifying code must satisfy the proposition
$$ P = \bigwedge_i C_i \bigwedge_{i \neq j} (C_i \bigoplus C_j) $$

For the de Bruijn graph $\mathcal{B}(2,4)$, we can compute $P$ explicitly and simplify the CNF to obtain
\begin{align*}
P =&  ( x_2 \vee x_3 ) \wedge ( x_0 \vee x_2 \vee x_4 \vee x_5 \vee x_8 \vee x_9 ) \wedge ( x_0 \vee x_3 \vee x_6 \vee x_7 \vee x_8 \vee x_9 )\\
&\wedge ( x_0 \vee x_1 \vee x_2 \vee x_4 \vee x_9 \vee x_{10} ) \wedge ( x_4 \vee x_{12} ) \wedge ( x_0 \vee x_1 \vee x_6 \vee x_9 \vee x_{12} \vee x_{14} )\\
&\wedge ( x_0 \vee x_3 \vee x_4 \vee x_5 \vee x_8 \vee x_9 ) \wedge ( x_0 \vee x_2 \vee x_6 \vee x_7 \vee x_8 \vee x_9 )\\
&\wedge ( x_0 \vee x_1 \vee x_3 \vee x_4 \vee x_9 \vee x_{10} ) \wedge ( x_1 \vee x_5 \vee x_8 \vee x_{10} )\\
&\wedge ( x_1 \vee x_4 \vee x_9 \vee x_{10} \vee x_{11} ) \wedge ( x_0 \vee x_2 \vee x_5 \vee x_8 \vee x_9 \vee x_{12} )\\
&\wedge ( x_1 \vee x_3 \vee x_5 \vee x_{12} ) \wedge ( x_1 \vee x_2 \vee x_9 \vee x_{10} \vee x_{13} )\\
&\wedge ( x_1 \vee x_6 \vee x_9 \vee x_{11} \vee x_{14} \vee x_{15} ) \wedge ( x_1 \vee x_3 \vee x_7 \vee x_8 \vee x_{12} \vee x_{14} )\\
&\wedge ( x_1 \vee x_3 \vee x_6 \vee x_9 \vee x_{14} \vee x_{15} ) \wedge ( x_4 \vee x_5 \vee x_8 \vee x_9 \vee x_{11} )\\
&\wedge ( x_0 \vee x_1 \vee x_2 \vee x_9 \vee x_{10} \vee x_{12} ) \wedge ( x_3 \vee x_8 \vee x_{10} \vee x_{12} )\\
&\wedge ( x_2 \vee x_5 \vee x_8 \vee x_9 \vee x_{13} ) \wedge ( x_2 \vee x_4 \vee x_{11} \vee x_{13} )\\
&\wedge ( x_2 \vee x_6 \vee x_7 \vee x_{10} \vee x_{13} ) \wedge ( x_2 \vee x_5 \vee x_6 \vee x_{13} \vee x_{14} )\\
&\wedge ( x_3 \vee x_5 \vee x_7 \vee x_{12} ) \wedge ( x_3 \vee x_{10} \vee x_{12} \vee x_{14} )\\
&\wedge ( x_3 \vee x_5 \vee x_6 \vee x_{13} \vee x_{14} \vee x_{15} ) \wedge ( x_3 \vee x_6 \vee x_7 \vee x_{10} \vee x_{13} \vee x_{15} )\\
&\wedge ( x_3 \vee x_{11} ) \wedge ( x_0 \vee x_1 \vee x_4 \vee x_6 \vee x_9 \vee x_{14} ) \wedge ( x_4 \vee x_6 \vee x_7 \vee x_{10} \vee x_{11} )\\
&\wedge ( x_4 \vee x_5 \vee x_6 \vee x_{11} \vee x_{14} ) \wedge ( x_5 \vee x_7 \vee x_{10} \vee x_{14} )\\
&\wedge ( x_5 \vee x_6 \vee x_{11} \vee x_{12} \vee x_{14} \vee x_{15} ) \wedge ( x_5 \vee x_6 \vee x_{11} \vee x_{13} \vee x_{14} \vee x_{15} )\\
&\wedge ( x_6 \vee x_7 \vee x_8 \vee x_9 \vee x_{13} \vee x_{15} ) \wedge ( x_6 \vee x_7 \vee x_8 \vee x_9 \vee x_{12} \vee x_{15} )\\
&\wedge ( x_6 \vee x_7 \vee x_{10} \vee x_{11} \vee x_{12} \vee x_{15} ) \wedge ( x_6 \vee x_7 \vee x_{10} \vee x_{11} \vee x_{13} \vee x_{15} )\\
&\wedge ( x_{12} \vee x_{13} ) \wedge ( x_0 \vee x_1 \vee x_8 ) \wedge ( x_1 \vee x_2 \vee x_4 \vee x_5 \vee x_9 )\\
&\wedge ( x_1 \vee x_3 \vee x_6 \vee x_7 \vee x_9 ) \wedge ( x_2 \vee x_4 \vee x_8 \vee x_9 \vee x_{10} )\\
&\wedge ( x_2 \vee x_5 \vee x_{10} \vee x_{11} ) \wedge ( x_4 \vee x_5 \vee x_{10} \vee x_{13} )\\
&\wedge ( x_5 \vee x_6 \vee x_7 \vee x_{11} \vee x_{13} ) \wedge ( x_6 \vee x_8 \vee x_9 \vee x_{12} \vee x_{14} )\\
&\wedge ( x_6 \vee x_{10} \vee x_{11} \vee x_{13} \vee x_{14} ) \wedge ( x_7 \vee x_{14} \vee x_{15} )
\end{align*}

Unfortunately, the above formula, consisting of 50 clauses over 16 variables, was slightly too large to map onto the architecture of the D-Wave processor we were using for this study.
We can decompose the problem into smaller subproblems by observing that $P$ contains the clauses $x_3 \vee x_{11}$ and $x_4 \vee x_{12}$.
Hence if $P$ is satisfied, at least one of $x_3$ and $x_{11}$ must be true, and similarly at least one of $x_4$ and $x_{12}$ must be true.
So we consider the following 4 cases:

\begin{align*}
1) \quad & x_3 = x_4 = 1 \\
2) \quad & x_3 = x_{12} = 1 \\
3) \quad & x_{11} = x_4 = 1 \\
4) \quad & x_{11} = x_{12} = 1
\end{align*}

While these 4 cases are not disjoint, if we find the minimum identifying codes for each case and take the union of the solutions, then the minimum length codes over the union will be the solutions of the original minimum identifying code problem.

We will illustrate the procedure for case 1.
If we set $x_3 = x_4 = 1$ in $P$, we obtain the reduced formula (24 clauses over 14 variables)
\begin{align*}
P' = & ( x_0 \vee x_1 \vee x_6 \vee x_9 \vee x_{12} \vee x_{14} ) \wedge  ( x_0 \vee x_2 \vee x_6 \vee x_7 \vee x_8 \vee x_9 ) \wedge\\
&( x_1 \vee x_5 \vee x_8 \vee x_{10} ) \wedge ( x_0 \vee x_2 \vee x_5 \vee x_8 \vee x_9 \vee x_{12} ) \wedge\\
&( x_1 \vee x_2 \vee x_9 \vee x_{10} \vee x_{13} ) \wedge( x_1 \vee x_6 \vee x_9 \vee x_{11} \vee x_{14} \vee x_{15} ) \wedge\\
&( x_0 \vee x_1 \vee x_2 \vee x_9 \vee x_{10} \vee x_{12} ) \wedge( x_2 \vee x_5 \vee x_8 \vee x_9 \vee x_{13} ) \wedge\\
&( x_2 \vee x_6 \vee x_7 \vee x_{10} \vee x_{13} ) \wedge( x_2 \vee x_5 \vee x_6 \vee x_{13} \vee x_{14} ) \wedge\\
&( x_5 \vee x_7 \vee x_{10} \vee x_{14} ) \wedge( x_5 \vee x_6 \vee x_{11} \vee x_{12} \vee x_{14} \vee x_{15} ) \wedge\\
&( x_5 \vee x_6 \vee x_{11} \vee x_{13} \vee x_{14} \vee x_{15} ) \wedge( x_6 \vee x_7 \vee x_8 \vee x_9 \vee x_{13} \vee x_{15} ) \wedge\\
&( x_6 \vee x_7 \vee x_8 \vee x_9 \vee x_{12} \vee x_{15} ) \wedge( x_6 \vee x_7 \vee x_{10} \vee x_{11} \vee x_{12} \vee x_{15} ) \wedge\\
&( x_6 \vee x_7 \vee x_{10} \vee x_{11} \vee x_{13} \vee x_{15} ) \wedge( x_{12} \vee x_{13} ) \wedge( x_0 \vee x_1 \vee x_8 ) \wedge\\
&( x_2 \vee x_5 \vee x_{10} \vee x_{11} ) \wedge( x_5 \vee x_6 \vee x_7 \vee x_{11} \vee x_{13} ) \wedge\\
&( x_6 \vee x_8 \vee x_9 \vee x_{12} \vee x_{14} ) \wedge( x_6 \vee x_{10} \vee x_{11} \vee x_{13} \vee x_{14} ) \wedge ( x_7 \vee x_{14} \vee x_{15} )
\end{align*}
	
This formula is small enough to map onto the D-Wave processor, as will be described in the following sections.

\subsection{Mapping SAT Clauses to Ising Models}\label{subsec:SAT2Ising}

The reduced proposition $P'$ for an identifying code has the CNF form
$$ P' = \bigwedge_j \bigvee_{i \in A_j} x_i$$
where each $A_j$ is a subset of the vertices of the de Bruijn graph.

To map this to an Ising model, we define ``spin'' variables $S_i = \pm1$ for each $i$:
$$ S_i = 2x_i - 1$$
We will also define as needed, some ``ancillary'' variables $z_k = \pm1$.

We construct an Ising Hamiltonian of the form
$$\mathcal{H} ( \{S_i\}, {z_k} ) = \sum_j \mathcal{H}_j + \lambda \sum_i S_i.$$
The Hamiltonian is a function of the problem variables $\{S_i\}$ and ancillary variables $\{z_k\}$.
Each of the terms $\mathcal{H}_j$ will be a function of the $\{S_i : i \in A_j \}$, and possibly some of the ancillary variables $\{z_k\}$, with the following properties:
\begin{list}{$\bullet$}{}
	\item $\mathcal{H}_j$ is at most quadratic in $\{S_i : i \in A_j \}$, and $\{z_k\}$
	\item $\{S_i\}$ is a minimum of $\mathcal{H}_j$ iff $\bigvee_{i \in A_j} x_i = 1$.
\end{list}

We will show momentarily how the $\mathcal{H}_j$ are constructed.  The last term $\lambda > 0$ is a penalty term that rewards smaller size codes.  Therefore, the minimum solutions (or ground states) of $\mathcal{H}$ are the minimum identifying codes.

To illustrate how the $\mathcal{H}_j$ are constructed, consider the 3-OR clause $x_1 \vee x_2 \vee x_3$.
If we define

$$	\mathcal{H}_3(S_1,S_2,S_3,z_1) = S_1 S_2 - 2S_1 z_1 - 2S_2 z_1 - 2S_2 z_1 + z_1 S_3 + S_1 + S_2 - 3z_1 - S_3$$

then it can be easily checked that $\mathcal{H}_3$ attains its minimum value iff at least one of the $S_i = +1$, which corresponds to the clause $x_1 \vee x_2 \vee x_3$ being satisfied.
Note that $\mathcal{H}_3$ contains no higher than quadratic terms.

\begin{figure}
	\begin{tikzpicture}[-,>=stealth',auto,node distance=2cm,
	thick,main node/.style={circle,draw,font=\sffamily\bfseries},new node/.style={font=\sffamily\bfseries},bend angle = 15]
	
	\node[main node] (21)                 {$S_1$};
	\node[new node,node distance=6mm] (21L) [above of=21] {$-1$};
	\node[main node] (22) [right of=21]   {$S_2$};
	\node[new node,node distance=6mm] (22L) [above of=22] {$-1$};
	\node[new node] (2) [above right of=21]   {2-OR};
	
	\node[main node] (31) [right of=22]   {$S_1$};
	\node[new node,node distance=6mm] (31L) [above of=31] {$+1$};
	\node[main node] (32) [right of=31]   {$S_2$};
	\node[new node,node distance=6mm] (32L) [above of=32] {$+1$};
	\node[main node,node distance=15mm] (3z1)[below right of=31] {$z_1$};
	\node[new node,node distance=7mm] (3z1L) [left of=3z1] {$-3$};
	\node[main node] (33) [right of=3z1] {$S_3$};
	\node[new node,node distance=6mm] (33L) [above of=33] {$-1$};
	\node[new node] (3) [above right of=31]  {3-OR};
	
	\node[new node]     (4)  [below of=22] {4-OR};
	\node[main node]    (41) [below left of=4]  {$S_1$};
	\node[new node,node distance=6mm] (41L) [above of=41] {$+1$};
	\node[main node]    (42) [right of=41]  {$S_2$};
	\node[new node,node distance=6mm] (42L) [above of=42] {$+1$};
	\node[main node,node distance=15mm]    (4z1)[below left of=42] {$z_1$};
	\node[new node,node distance=7mm] (4z1L) [left of=4z1] {$-3$};
	\node[main node]    (43) [right of=42]  {$S_3$};
	\node[new node,node distance=6mm] (43L) [above of=43] {$+1$};
	\node[main node]    (44) [right of=43]  {$S_4$};
	\node[new node,node distance=6mm] (44L) [above of=44] {$+1$};
	\node[main node,node distance=15mm]    (4z2)[below right of=43] {$z_2$};
	\node[new node,node distance=6mm] (4z2L) [right of=4z2] {$-3$};

	\node[new node]     (5)  [below of=4z2]  {5-OR};
	\node[main node,node distance=15mm]    (52) [below left of=5] {$S_2$};
	\node[new node,node distance=6mm] (52L) [above of=52] {$+1$};
	\node[main node,node distance=15mm]    (53) [below right of=5] {$S_3$};
	\node[new node,node distance=6mm] (53L) [above of=53] {$+1$};
	\node[main node]    (51) [left of=52]       {$S_1$};
	\node[new node,node distance=6mm] (51L) [above of=51] {$+1$};
	\node[main node]    (54) [right of=53]      {$S_4$};
	\node[new node,node distance=6mm] (54L) [above of=54] {$+1$};
	\node[main node,node distance=15mm]    (5z1)[below left of=52] {$z_1$};
	\node[new node,node distance=7mm] (5z1L) [left of=5z1] {$-1$};
	\node[main node,node distance=15mm]    (5z2)[below right of=53]{$z_2$};
	\node[new node,node distance=6mm] (5z2L) [right of=5z2] {$-1$};
	\node[main node,node distance=30mm]    (5z3)[below right of=5z1]{$z_3$};
	\node[new node,node distance=7mm] (5z3L) [left of=5z3] {$-3$};
	\node[main node]    (55) [right of=5z3]     {$S_5$};
	\node[new node,node distance=6mm] (55L) [above of=55] {$-1$};
	
	\node[new node]     (6) [below of=5z3]    {6-OR};
	\node[main node,node distance=15mm] (62)    [below left of=6]   {$S_2$};
	\node[new node,node distance=6mm]   (62L)   [above of=62]       {$+1$};
	\node[main node,node distance=15mm] (63)    [below right of=6]  {$S_3$};
	\node[new node,node distance=6mm]   (63L)   [above of=63]       {$+1$};
	\node[main node]                    (61)    [left of=62]        {$S_1$};
	\node[new node,node distance=6mm]   (61L)   [above of=61]       {$+1$};
	\node[main node]                    (64)    [right of=63]       {$S_4$};
	\node[new node,node distance=6mm]   (64L)   [above of=64]       {$+1$};
	\node[main node,node distance=15mm] (6z1)   [below left of=62]  {$z_1$};
	\node[new node,node distance=7mm]   (6z1L)  [left of=6z1]       {$-1$};
	\node[main node,node distance=15mm] (6z2)   [below right of=63] {$z_2$};
	\node[new node,node distance=6mm]   (6z2L)  [right of=6z2]      {$-1$};
	\node[main node,node distance=30mm] (6z3)   [below right of=6z1]{$z_3$};
	\node[new node,node distance=7mm]   (6z3L)  [left of=6z3]       {$-3$};
	\node[main node,node distance=45mm] (6z4)   [right of=6z3]      {$z_4$};
	\node[new node,node distance=6mm]   (6z4L)  [right of=6z4]      {$-3$};
	\node[main node,node distance=15mm] (65)    [above left of=6z4] {$S_5$};
	\node[new node,node distance=6mm]   (65L)   [below of=65]       {$+1$};
	\node[main node,node distance=15mm] (66)    [above right of=6z4]{$S_6$};
	\node[new node,node distance=6mm]   (66L)   [below of=66]       {$+1$};
	
	\path[every node/.style={font=\sffamily\footnotesize}]
	
	(21) edge node [above]      {+1} (22)
	
	(31)edge node [above]       {+1} (32)
	edge node [left]        {-2} (3z1)
	(32)edge node [right]       {-2} (3z1)
	(33)edge node [below]       {+1} (3z1)
	
	(41) edge node [above]      {+1} (42)
	(41) edge node [left]       {-2}    (4z1)
	(4z1) edge node [right]     {-2}    (42)
	(4z1) edge node [below]     {+1}    (4z2)
	(4z2) edge node [left]      {-2}    (43)
	(4z2) edge node [right]     {-2}    (44)
	(43) edge node [above]      {+1}    (44)
	
	(51) edge node [above]      {+1}    (52)
	(51) edge node [left]       {-2}    (5z1)
	(52) edge node [right]      {-2}    (5z1)
	(5z1) edge node [above]     {+1}    (5z2)
	(5z1) edge node [left]      {-2}    (5z3)
	(5z3) edge node [above]     {+1}    (55)
	(5z3) edge node [right]     {-2}    (5z2)
	(5z2) edge node [left]      {-2}    (53)
	(5z2) edge node [right]     {-2}    (54)
	(53) edge node [above]      {+1}    (54)
	
	(61)    edge    node    [above] {+1}    (62)
	(61)    edge    node    [left]  {-2}    (6z1)
	(62)    edge    node    [right] {-2}    (6z1)
	(63)    edge    node    [above] {+1}    (64)
	(63)    edge    node    [left]  {-2}    (6z2)
	(64)    edge    node    [right] {-2}    (6z2)
	(6z1)   edge    node    [above] {+1}    (6z2)
	(6z1)   edge    node    [left]  {-2}    (6z3)
	(6z2)   edge    node    [right] {-2}    (6z3)
	(6z3)   edge    node    [above] {+1}    (6z4)
	(66)    edge    node    [left]  {-2}    (6z4)
	(65)    edge    node    [above] {+1}    (66)
	(65)    edge    node    [right] {-2}    (6z4)
	;
	
	\end{tikzpicture}
	\caption{Mapping from OR-clauses to Ising Models}\label{fig:Gadgets}
\end{figure}
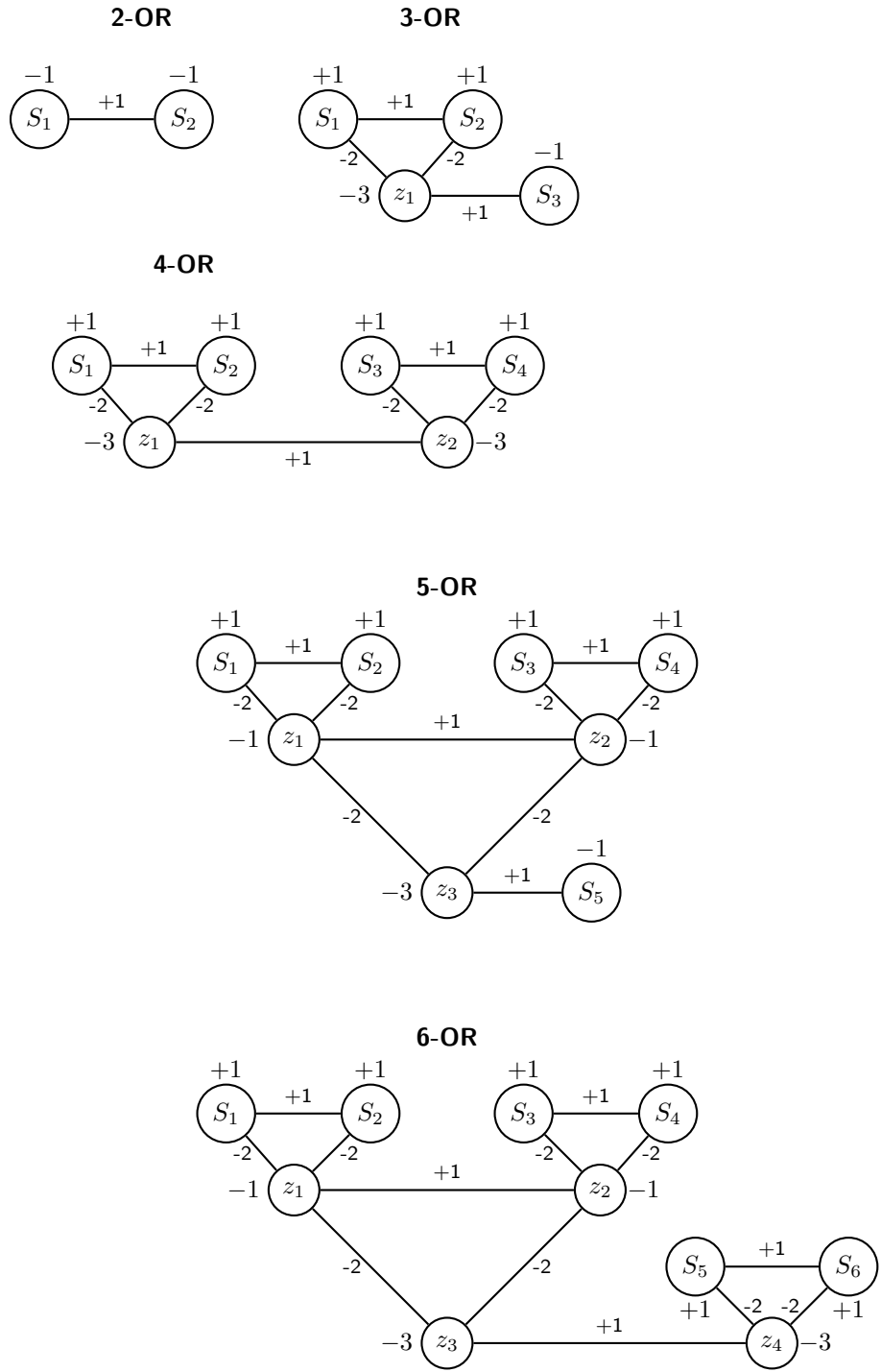

We refer to the mapping from the 3-OR clause to $\mathcal{H}_3$ as a ``gadget''.
The gadget can be represented diagrammatically as in Figure \ref{fig:Gadgets}, which also shows gadgets for 4-OR through 6-OR clauses, which are the gadgets we need to map our satisfiability formula $P'$ for $\mathcal{B}(d,n)$ to an Ising model.
In the diagrams, numbers attached to a node represent the linear coefficients in the Ising model, while numbers attached to an edge represent the quadratic (coupling) coefficients in the Ising model.

This technique is similar to the ones described in \cite{Gadgets}, except that those gadgets were designed for use with 0/1 variables instead of $\pm 1$ variables.
Note that the choice of gadget coefficients is far from unique, and it may be possible to tune these coefficients, for example to accommodate the limited control precision of the quantum processor.
However, the coefficients shown in Figure \ref{fig:Gadgets} were sufficient for the problem at hand.
When larger D-Wave processors become available that enable solving the minimum identifying code problem for larger $(d,n)$, tuning of the gadget coefficients may be needed, along with the gauge transformation techniques described in the next subsection.

Using the gadgets shown in Figure \ref{fig:Gadgets}, we mapped the satisfiability formula $P'$ to an Ising model with 49 ancillary variables $\{z_k\}$, for a total of 63 variables.
We furthermore added the penalty term $\lambda \sum_i S_i$ so that the ground state will be a minimum identifying code.
Since this Ising model could in principle be implemented on an ideal quantum annealing machine with 63 qubits, we say that the model has 63 \textit{logical qubits}, and we refer to this model as the \textit{logical Ising model}.

\begin{figure}[ht]
	\centering
	\includegraphics[width=0.8\textwidth]{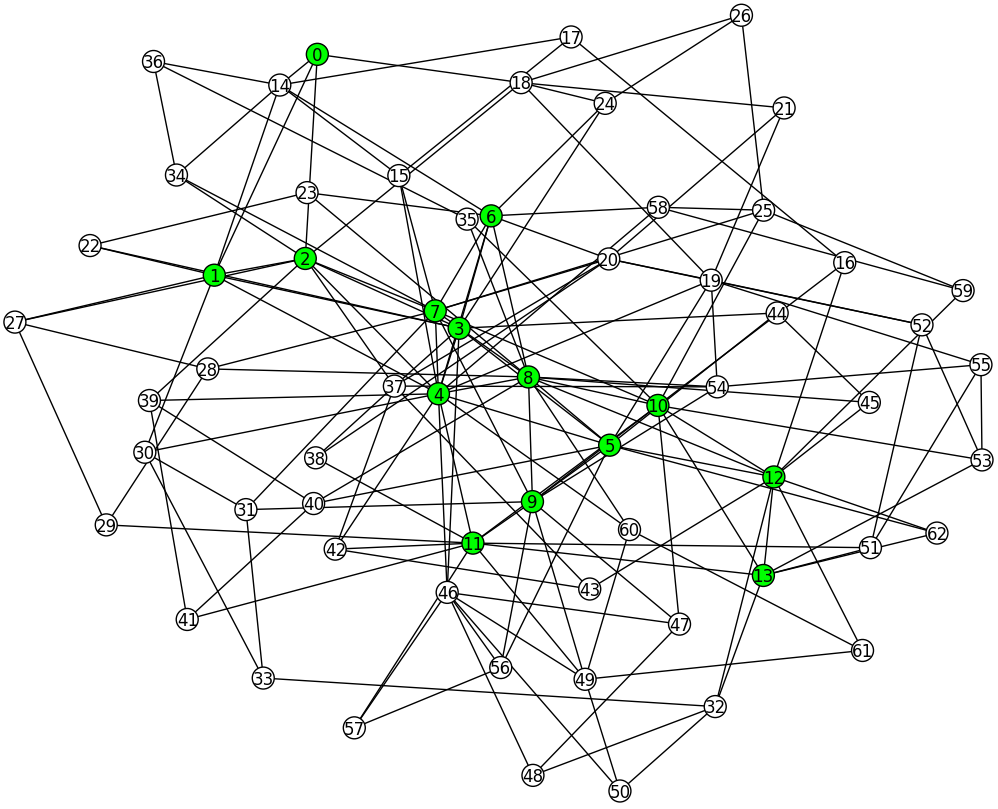}
	\caption{Graph of logical Ising model for the SAT formula $P'$.  Nodes corresponding to the original 14 boolean variables are shown in green; the remaining nodes represent the ancillary variables added during the SAT-to-Ising mapping process.}\label{fig:logicalGraph}
\end{figure}

While the overhead incurred in the satisfiability-based approach to obtain a quadratic Ising model is substantial (going from 14 boolean variables in $P'$ to 63 logical qubits in the Ising model), one advantage of this approach is that the connectivity of the resulting Ising model (i.e. pairs of variables with nonzero coefficients) is relatively sparse.
This can be seen from Figure \ref{fig:logicalGraph}.
In the figure, nodes corresponding to the original 14 boolean variables are shown in green; the remaining nodes represent the ancillary variables added during the SAT-to-Ising mapping process.
Edges represent pairs of variables with nonzero coefficients ($J_{ij} \neq 0$).
The relative sparsity of this graph will facilitate mapping the problem onto the D-Wave architecture, as described in the next subsection.

\newpage

\subsection{Mapping the Logical Ising Model onto the D-Wave Processor} \label{subsec:Ising2DWave}

This subsection describes additional pre-processing steps that take place before the logical Ising model is solved on the D-Wave processor.
\textit{Graph minor embedding} enables the logical Ising model to be mapped onto the limited qubit connectivity of the D-Wave architecture, by utilizing potentially multiple physical qubits per logical qubit.
To improve solution accuracy, \textit{gauge transformations} are used to partially mitigate the intrinsic control errors that occur when programming the D-Wave hardware.

\subsubsection{Graph Minor Embedding}

The physical connectivity between qubits on the D-Wave chip can be represented as a square lattice of $K_{4,4}$ bipartite graphs, that D-Wave calls a \textit{Chimera graph}.
Even for Ising models with relatively sparse graphs, it may not be possible to directly map the Ising model onto the Chimera graph.

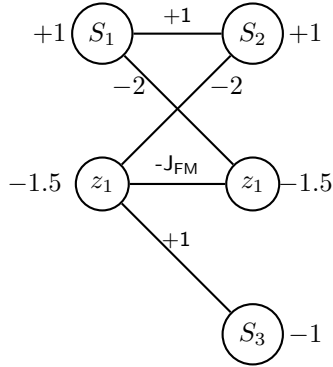
\begin{figure}
\begin{center}
\begin{tikzpicture}[-,>=stealth',auto,node distance=2cm,
  thick,main node/.style={circle,draw,font=\sffamily\bfseries},new node/.style={font=\sffamily\bfseries},bend angle = 15]

    \node[main node]    (1)                         {$S_1$};
    \node[new node,node distance=7mm]   (1L)        [left of=1]       {$+1$};
    \node[new node,node distance=10mm]  (12L)       [below right of=1,anchor=east]  {$-2$};
    \node[main node]    (2)     [right of=1]        {$S_2$};
    \node[new node,node distance=10mm]  (22L)       [below left of=2,anchor=west]  {$-2$};
    \node[new node,node distance=7mm]   (2L)        [right of=2]       {$+1$};
    \node[main node]    (z1)    [below of=1]        {$z_1$};
    \node[new node,node distance=9mm]   (z1L)        [left of=z1]       {$-1.5$};
    \node[main node]    (z2)    [right of=z1]       {$z_1$};
    \node[new node,node distance=7mm]   (z2L)        [right of=z2]       {$-1.5$};
    \node[main node]    (3)     [below of=z2]       {$S_3$};
    \node[new node,node distance=7mm]   (3L)        [right of=3]       {$-1$};

  \path[every node/.style={font=\sffamily\footnotesize}]

    (1) edge node [above]      {+1} (2)
    (1) edge node [above left,distance=2mm]  {}    (z2)
    (2) edge node [above right] {}    (z1)
    (z1) edge node [above]      {$\textsf{-J}_{\textsf{FM}}$}  (z2)
    (z1) edge node [above]      {+1}    (3)
    ;
\end{tikzpicture}
\end{center}
\caption{Embedding 3-OR onto the Chimera Graph}\label{fig:3-ORembedding}
\end{figure}

For example, even the simple Ising model for 3-OR (the $\mathcal{H}_3$ gadget) shown in Figure \ref{fig:Gadgets} cannot be directly mapped onto the Chimera graph.
An easy way to see this is to observe that the graph of $\mathcal{H}_3$ contains a 3-cycle, whereas the smallest cycle possible on the Chimera graph is a 4-cycle.
However, the graph of $\mathcal{H}_3$ can be mapped onto the Chimera graph using minor embedding.
One such embedding is shown in Figure \ref{fig:3-ORembedding}.
In the figure, the logical qubit $z_1$ is mapped to two physical qubits, which are ferromagnetically coupled with a coupling strength $\textsf{-J}_{\textsf{FM}}$.
The $h$ and $J$ values shown in the figure define an embedded Hamiltonian which only uses connections that are available in the Chimera graph.
It is easily verified that the ground state of the embedded Hamiltonian corresponds to the ground state of the original Hamiltonian $\mathcal{H}_3$.

Choi \cite{Choi2} showed that any graph can be minor embedded into a sufficiently large Chimera graph.
Heuristic embedding algorithms have been developed \cite{Cai} to generate minor embeddings for Chimera graphs with faulty qubits.
The D-Wave software includes a heuristic embedding tool that can be used to find embeddings using the working qubits on an actual D-Wave chip.

\begin{figure}[p]
	\begin{center}
		\includegraphics[width=0.8\textwidth]{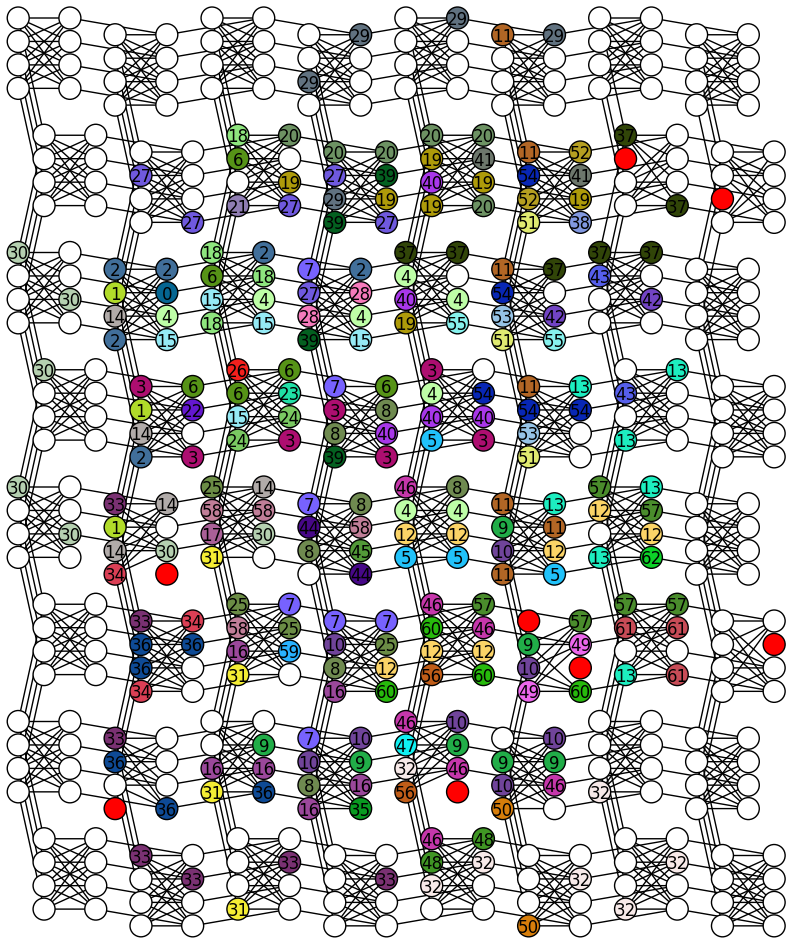}
	\end{center}
	\vspace{-15pt}
	\caption{Embedding the problem onto the D-Wave hardware:  physical qubits corresponding to the same logical qubit have the same color and are labeled with the same number; the unlabeled red qubits are known faulty qubits and are not used.}\label{fig:ChimeraEmbedding}
\end{figure}

Using the D-Wave heuristic embedding tool, we mapped the graph of the logical Ising model, that was shown in Figure \ref{fig:logicalGraph}, into the hardware graph of our D-Wave processor which had 504 working qubits.
Figure \ref{fig:ChimeraEmbedding} shows the embedding that we used, consisting of 253 physical qubits, with a maximum chain length (number of physical qubits per logical qubit) of 8.
The coefficients of the physical Ising model are determined as in \cite{Choi1}.

\subsubsection{Gauge Transformations}

In the context of quantum annealing, a \textit{gauge transformation} \cite{Boixo} is a transformation of the Ising spin variables
$$ S_i' = G_i S_i $$
where $G_i \in \{-1,+1\}$.
The gauge transformation induces a transformation on the physical Ising model coefficients
\begin{align*}
h_i' &= G_i h_i \\
J_{ij}' &= G_i G_j J_{ij}
\end{align*}
so that the Hamiltonian is invariant.

While the gauge transformation generates a problem that is mathematically equivalent to the original problem, in practice it has been found that gauge transformations can  mitigate some of the intrinsic control error (ICE) of the D-Wave hardware \cite{King} and that the choice of the gauge $G$ can affect the probability of finding optimal solutions \cite{Perdomo-Ortiz}.

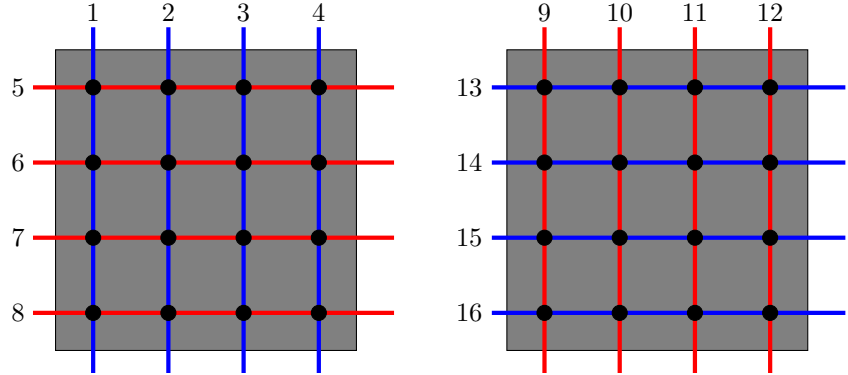
\begin{figure}

\begin{center}
    \begin{tikzpicture}
        \node (8) at (0,1) {8};
        \node (7) at (0,2) {7};
        \node (6) at (0,3) {6};
        \node (5) at (0,4) {5};
        \node (1) at (1,5) {1};
        \node (2) at (2,5) {2};
        \node (3) at (3,5) {3};
        \node (4) at (4,5) {4};
        \draw[fill=gray] (0.5,0.5) rectangle (4.5,4.5);
        \draw[ultra thick,draw=red] (0.2,4) -- (5,4);
        \draw[ultra thick,draw=red] (0.2,3) -- (5,3);
        \draw[ultra thick,draw=red] (0.2,2) -- (5,2);
        \draw[ultra thick,draw=red] (0.2,1) -- (5,1);
        \draw[ultra thick,draw=blue](1,4.8) -- (1,0.2);
        \draw[ultra thick,draw=blue](2,4.8) -- (2,0.2);
        \draw[ultra thick,draw=blue](3,4.8) -- (3,0.2);
        \draw[ultra thick,draw=blue](4,4.8) -- (4,0.2);
        \draw[fill=black] (4,1) circle (1mm);
        \draw[fill=black] (4,2) circle (1mm);
        \draw[fill=black] (4,3) circle (1mm);
        \draw[fill=black] (4,4) circle (1mm);
        \draw[fill=black] (3,1) circle (1mm);
        \draw[fill=black] (3,2) circle (1mm);
        \draw[fill=black] (3,3) circle (1mm);
        \draw[fill=black] (3,4) circle (1mm);
        \draw[fill=black] (2,1) circle (1mm);
        \draw[fill=black] (2,2) circle (1mm);
        \draw[fill=black] (2,3) circle (1mm);
        \draw[fill=black] (2,4) circle (1mm);
        \draw[fill=black] (1,1) circle (1mm);
        \draw[fill=black] (1,2) circle (1mm);
        \draw[fill=black] (1,3) circle (1mm);
        \draw[fill=black] (1,4) circle (1mm);

        \node (16) at (6,1) {16};
        \node (15) at (6,2) {15};
        \node (14) at (6,3) {14};
        \node (13) at (6,4) {13};
        \node (9)  at (7,5) {9};
        \node (10) at (8,5) {10};
        \node (11) at (9,5) {11};
        \node (12) at (10,5) {12};
        \draw[fill=gray] (6.5,0.5) rectangle (10.5,4.5);
        \draw[ultra thick,draw=blue] (6.3,4) -- (11,4);
        \draw[ultra thick,draw=blue] (6.3,3) -- (11,3);
        \draw[ultra thick,draw=blue] (6.3,2) -- (11,2);
        \draw[ultra thick,draw=blue] (6.3,1) -- (11,1);
        \draw[ultra thick,draw=red](7,4.8) -- (7,0.2);
        \draw[ultra thick,draw=red](8,4.8) -- (8,0.2);
        \draw[ultra thick,draw=red](9,4.8) -- (9,0.2);
        \draw[ultra thick,draw=red](10,4.8) -- (10,0.2);
        \draw[fill=black] (7,1) circle (1mm);
        \draw[fill=black] (7,2) circle (1mm);
        \draw[fill=black] (7,3) circle (1mm);
        \draw[fill=black] (7,4) circle (1mm);
        \draw[fill=black] (8,1) circle (1mm);
        \draw[fill=black] (8,2) circle (1mm);
        \draw[fill=black] (8,3) circle (1mm);
        \draw[fill=black] (8,4) circle (1mm);
        \draw[fill=black] (9,1) circle (1mm);
        \draw[fill=black] (9,2) circle (1mm);
        \draw[fill=black] (9,3) circle (1mm);
        \draw[fill=black] (9,4) circle (1mm);
        \draw[fill=black] (10,1) circle (1mm);
        \draw[fill=black] (10,2) circle (1mm);
        \draw[fill=black] (10,3) circle (1mm);
        \draw[fill=black] (10,4) circle (1mm);
    \end{tikzpicture}
\end{center}

\caption{Example Gauge Transformation}\label{fig:GT}
\end{figure}

Consider for example the gauge transformation $G$ shown in Figure \ref{fig:GT}.
The figure depicts the action of $G$ on the first unit cell, where the red qubits are flipped by $G$ while the blue qubits are unchanged;
i.e. all of the horizontal qubits are flipped while the vertical qubits are unchanged.
In the next unit cell, all of the vertical qubits are flipped while the horizontal qubits are unchanged; and so on in an alternating pattern.

We have found that this gauge transformation $G$ is particularly helpful in connection with embedding.
As mentioned earlier, the embedding process maps a single logical qubit to multiple physical qubits that are ``chained'' together with a ferromagnetic coupling $\textsf{-J}_{\textsf{FM}}$.
Furthermore, to enforce the embedding constraints (i.e. that all the physical qubits in a chain should take the same value), the coupling strength $\textsf{J}_{\textsf{FM}}$ is generally made to be the dominant coupling in the Hamiltonian.
The gauge transformation $G$ has the property that all the ferromagnetic couplings $\textsf{-J}_{\textsf{FM}}$ in each chain are replaced by antiferromagnetic couplings $\textsf{+J}_{\textsf{FM}}$.
This can help to mitigate certain types of systematic ICE errors that are magnified by long ferromagnetically coupled chains.

\subsection{Solution for $\mathcal{B}(2,4)$ Using D-Wave Processor}\label{subsec:SolnUsingDWave}

Using the reduced proposition $P'$, mapped to a logical Ising model using the gadgets shown in Figure \ref{fig:Gadgets}, and the embedding shown in Figure \ref{fig:ChimeraEmbedding}, we used the 504-qubit LM/USC D-Wave processor to solve the resulting physical Ising model using quantum annealing.

Since the minimum identifying code problem for $\mathcal{B}(2,4)$ is small enough to solve by brute force, we know that the reduced proposition $P'$ should be satisfied by $$x_8=x_{10}=x_{13}=x_{14}=1$$ (and the rest of the variables zero).
Combined with the assumption $x_3 = x_4 =1$, this correspond to a minimum code size of 6.

For the non-gauge transformed problem, we found that the D-Wave machine did not obtain the optimal solution.
In 480 experiments totaling 630,000 annealing runs, over a range of different settings for the annealing time, we obtained zero occurrences of the optimal solution; the best that we found were solutions corresponding to a code size of 7.

On the other hand, using the gauge transformation $G$ shown in Figure \ref{fig:GT}, in 240 experiments totaling 825,000 annealing runs (again using a range of settings for the annealing time), we obtained the solution $$x_8=x_{10}=x_{13}=x_{14}=1$$ corresponding to the minimum code size of 6 in 25 experiments.
So, while the ground state probability was still very low in the gauge transformed problem, there was a noticeable difference using the gauge transformation.

\subsection{Scaling for Larger Cases}\label{subsec:Scaling}

\begin{figure}[p]
\begin{center}
\includegraphics[width=\textwidth]{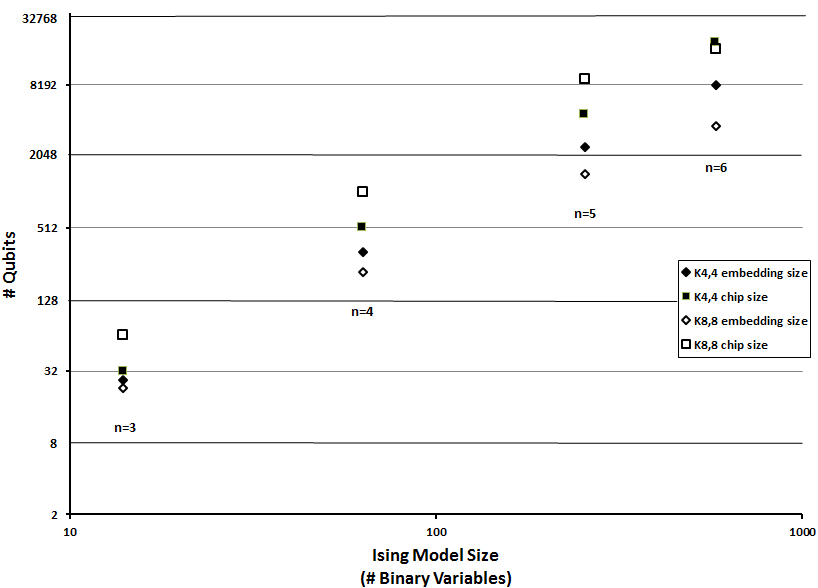}
\end{center}
\vspace{-15pt}
\caption{Estimated scaling of qubit resource requirements for minimum identifying code problem on binary de Bruijn graphs  $\mathcal{B}(2,n)$}\label{QubitScaling}
\end{figure}

While the $\mathcal{B}(2,4)$ case was the largest that could be solved on the 504-qubit LM/USC D-Wave processor, we can make some rough estimates of how the qubit resource requirements scale for larger cases.
For binary ($d=2$) de Bruijn graphs, we extended the techniques described above to generate Ising models for the minimum identifying code problem for the cases $n=3$ through $n=6$.
These models were then embedded into ideal Chimera graphs of various sizes using the D-Wave heuristic embedding tool.
Figure \ref{QubitScaling} shows how the number of qubits needed for the embedding grows as a function of $n$.

Since the embedding algorithm is heuristic, it is possible that smaller embeddings could be found, so the embedding sizes shown here should be viewed as upper bounds.
We computed embeddings firstly for larger versions of the current $K_{4,4}$ Chimera architecture, which has 8-qubit unit cells, as well as for  hypothetical $K_{8,8}$ (16-qubit unit cell) Chimera graphs.
For the $K_{4,4}$ Chimera graphs, we started with the current Vesuvius architecture, which contains 64 8-qubit unit cells arranged in an 8x8 square grid, and increased the grid size to 12x12, 16x16, 24x24, 32x32, and so on until the embedding was successful.
For the $K_{8,8}$ graphs, we started with an 8x8 grid of 16-qubit unit cells, and increased the grid size to 16x16,24x24, and 32x32, until the embedding was successful.
The plots show that for any given case, the embedding size on the $K_{8,8}$ Chimera is smaller than on the $K_{4,4}$ Chimera due to the higher graph connectivity of the $K_{8,8}$ Chimera architecture; in other words, qubit resource requirements depend on the hardware graph connectivity.

From the figure, we can project roughly when the D-Wave processor would have sufficient qubits to accommodate the larger cases of the minimum identifying code problem for undirected binary de Bruijn graphs.
If we define a processor generation to be 4 times the number of qubits as the previous generation (e.g. the ``Vesuvius'' generation had a 512-qubit design whereas the prior ``Rainier'' generation had a 128-qubit design), we can state that in roughly 1.5 generations we would be able to fit the $\mathcal{B}(2,5)$ case on the processor.
This is the largest $d=2$ case for which the minimum identifying code size is presently known.
In one more generation beyond that, we would be able to fit the $\mathcal{B}(2,6)$ case on the processor, for which the minimum identifying code size is presently not known.

Whether these hypothetical D-Wave processors would actually be able to solve these larger cases, will depend on the performance characteristics of those machines which is yet to be demonstrated.
The hardware intrinsic control errors would need to be significantly reduced from the current levels.
For example, the embedding we found for the $\mathcal{B}(2,6)$ case on the $K_{4,4}$ Chimera architecture had a maximum chain length of 63 qubits.
Solving a problem with chain lengths this large would probably require greater control precision (the current Vesuvius design has 4 bits of precision \cite{Gadgets}).
On the $K_{8,8}$ Chimera architecture, the embeddings are less complex; e.g. the embedding we found for the $\mathcal{B}(2,6)$ case only had a maximum chain length of 25 qubits.
However, it is not clear how difficult it would be for D-Wave to achieve a 16-qubit unit cell design.

Due to risk factors such as these, it could take longer than the projected number of processor generations before a solution can be found to the $\mathcal{B}(2,6)$ case.
On the other hand, it may be possible to break the problem down into subproblems, as illustrated earlier in Subsection \ref{subsec:SATForm}, which may make it easier to fit the problem onto the D-Wave processor.

\subsection{Other Problem Formulations Considered}
\label{subsec:OtherForms}

While the satisfiability formulation provided the best result, we include two other methods here for the sake of completeness.  It should be noted that while these methods were not practical for our sample problem, they may provide advantages for other problems.

\subsubsection{Integer Programming Model}

From \cite{ICDG}, we have the following integer program formulation of the minimum identifying code problem in graphs.

First, we define the modified adjacency matrix as follows.  It is the adjacency matrix plus the identity matrix.

$$A_{ij} = \left\{
      \begin{array}{ll}
           1, & \hbox{if $(i,j) \in E$ or $i=j$;} \\
           0, & \hbox{otherwise.}
      \end{array}
\right.$$

Using this definition, we see that a ball of radius 1 centered at vertex $i$ is given by the following vector.  $$B(i) = [A_{1i}, A_{2i}, \ldots , A_{ni}]^T$$  Our vertex subset $S$ is defined as the following vector.

$$S = [s_1, s_2, \ldots , s_n]^T \hbox{ where } s_i = \left\{ \begin{array}{ll} 1, & \hbox{if $i \in S$;} \\ 0, & \hbox{otherwise.}\end{array}\right.$$

To compare two identifying sets with respect to $S$ for vertices $i$ and $j$, the following expression computes the size of $(B(i) \cap S) \triangle (B(j) \cap S)$.  $$\sum_{k=1}^n |A_{ki}-A_{kj}| \cdot s_k$$

This implies that in order for $S$ to be a valid identifying code, we must have the following inequality satisfied for all pairs of vertices $i$ and $j$.  $$\sum_{k=1}^n |A_{ki}-A_{kj}| \cdot s_k \geq 1$$

For the dominating property to be satisfied, we require the following additional inequality.  $$A \cdot S \geq \mathbf{1}^T$$

Thus our integer program is given by the following.
$$\begin{array}{rlcll}
    \min & |S| \\
    \hbox{s.t.} & \sum_{k=1}^n |A_{ki}-A_{kj}| \cdot s_k & \geq & 1, & \forall i \neq j \\
    & A \cdot S & \geq & \mathbf{1}^T \\
    & s_k \in \{0,1\}
\end{array}$$

In order to use these ideas for the D-Wave machine, our constraints must be equalities.  This means we must add binary slack variables for each inequality.  For the first set of inequalities, we must determine an upper bound for each inequality.  Since these correspond to the constraint $|(B(i) \cap S) \triangle (B(j) \cap S) | \geq 1$, an easy upper bound is given by the following.  $$|B(i)| + |B(j)| \geq |(B(i) \cap S) \triangle (B(j) \cap S)| \geq 1$$

For the class of de Bruijn graphs, we are able to use this to get a bound on the number of slack variables needed.  Since the maximum size of any ball in $\mathcal{B}(d,n)$ is $2d+1$, this gives us an upper bound of size $4d+2$ for this class of graphs.  Hence for each inequality in this set, we must add $4d+2$ binary slack variables to convert the inequality to an equality.  Using a variable reduction method from \cite{isingPaper}, we can further reduce this number of variables to $\log (4d+2)$.  Since there are $\frac{d^n(d^n-1)}{2}$ possible pairs $i,j$, this implies that we must add a huge number of binary slack variables, equal to the following expression, just to satisfy the first set of inequalities.  $$\frac{d^n(d^n-1)\log{(4d+2)}}{2} \hbox{ slack variables}$$  Even in the case of $\mathcal{B}(2,4)$, this means that we will need to add 320 variables to our list - an enormous number when compared to the graph size of 16 nodes.  Hence, this method is not going to be an efficient way to map our problem onto the D-Wave machine.

\subsubsection{Binary Optimization Model}

We present a binary optimization formula for the minimum identifying code problem.  Adjustments must be made to create a quadratic version (QUBO), making this approach impractical for large scale results.  We will define this model using three separate functions:  one to show that the set has the correct size, one to show that the set is dominating, and one to show that the set is separating (or identifying).

\vspace{5mm}

\noindent\textbf{Variable Definitions}

We will use the notation $B(v)$ for $v \in V(G)$, where $B(v) = N(v) \cup \{v\}$.  In other words, $B(v)$ is the set containing all vertices adjacent to $v$, plus $v$ itself.  This is referred to in graph theory as the ball of radius one centered at $v$.

We define the variables as follows.  $$x_{vi} = \left\{
                                                  \begin{array}{ll}
                                                    1, & \hbox{if $i \in B(v)$;} \\
                                                    0, & \hbox{otherwise.}
                                                  \end{array}
                                                \right.$$

\noindent\textbf{Set $S$ has size $k$}

We define the first function, $H_A$, as follows.

$$\begin{array}{ccccc}
    H_A & = & \left(k-\sum_v x_{vv} \right) \\
& = & 0 & \hbox{iff} & |S|=k.
\end{array}$$

\noindent\textbf{Set $S$ is a dominating set}

By definition, this is equal to $\forall v \in G$, $B(v) \cap S \neq \emptyset$.  This is equivalent to the following.

\begin{eqnarray*}
    \forall v \in G, B(v) \cap S \neq \emptyset & \leftrightarrow & (x_{vv} = 1) \vee \left(\sum_{uv \in E} x_{uv} \geq 1 \right) \\
    & \leftrightarrow & (1-x_{vv} = 0) \vee \neg \left(\sum_{uv \in E} x_{uv} = 0\right) \\
    & \leftrightarrow & (1-x_{vv} = 0) \vee \left( \prod_{uv \in E} (1-x_{uv}) = 0 \right)
\end{eqnarray*}

From this statement, we get the following equation for our second function.

$$\begin{array}{ccccc}
    H_B & = & \sum_v (1-x_{vv}) \cdot \left( \prod_{uv \in E} (1-x_{uv}) \right)
\end{array}$$

\noindent\textbf{Set $S$ is a separating set}

By definition, this is equal to $\forall x,y \in G$, $(B(x) \cap S) \triangle (B(y) \cap S) \neq \emptyset$.  This is equivalent to the following for a specific pair $x \neq y$.

\begin{eqnarray*}
    (B(x) \cap S) \triangle (B(y) \cap S) \neq \emptyset & \leftrightarrow & \exists v \in (B(x) \cap S) \triangle (B(y) \cap S) \\
    & \leftrightarrow & \exists v, (v \in B(x) \cap S) \oplus (v \in B(y) \cap S) \\
    & \leftrightarrow & \exists v, (x_{xv}=1) \oplus (x_{yv}=1) \\
    & \leftrightarrow & \exists v, (1-(x_{xv}+x_{yv})=0) \\
    & \leftrightarrow & \prod_{v} (1-x_{xv}-x_{yv})=0
\end{eqnarray*}

From this statement, we get the following equation for our third function, summed over all pairs $x,y$.

$$\begin{array}{ccccc}
    H_C & = & \sum_x \sum_{y \neq x} \prod_v (1-x_{xv}-x_{yv})^2
\end{array}$$

\noindent\textbf{The Binary Optimization Model}

From these three functions, our binary optimization model is the following.
$$\begin{array}{ccccc}
    H(S) & = & H_A(S) + H_B(S) + H_C(S) \\
    & = & 0 & \hbox{iff} & $S$ \hbox{ is an identifying code.}
\end{array}$$

Note that while this does provide a binary optimization model for our problem, it is not quadratic.  In order to convert $H(S)$ to a quadratic binary equation, each higher order term must be replaced with several new variables.  While this is possible, it is a time-consuming and arduous process that introduces many new variables.  Hence this approach is not the most efficient implementation.

\section{Other Approaches to Solving the Problem}

\subsection{Parallel Computing}\label{Parallel}

The traditional brute force approach to solving the minimum identifying code problem constructs all possible subsets from smallest to largest size, and checks whether or not the set is a valid identifying code.  Parallelizing our algorithm (implemented in Matlab using the Parallel Computing Toolbox) requires moving the construction of subsets inside the parallelized loop.  Because of the exponential increase in the number of subsets created, it is more efficient to generate each subset within the loop and discard it after the iteration than to store all $\binom{d^n}{k}$ $k$-subsets and traverse through the list.  This is done using a $k$-subset unranking algorithm.  Two of these algorithms (from \cite{algos}) are listed as Algorithms \ref{Unrank1} and \ref{Unrank2}.  These unranking functions allow us to completely parallelize the brute force algorithm, and the results obtained are listed in Figure \ref{HPCRes}.

\begin{algorithm}
\caption{Revolving Door Unranking Algorithm}\label{Unrank1}
\begin{algorithmic}[1]
\Procedure{RevDoor}{$r,k,n$}
\Comment{subset index, subset size, set size}
\State $x=n$
\For{$i=k:1$}
\While{$\binom{x}{i} >r$}
\State $x=x-1$
\EndWhile
\State $t_i=x+1$
\State $r = \binom{x+1}{i}-r-1$
\EndFor
\EndProcedure
\State \textbf{return} $T=(t_1,t_2, \ldots , t_k)$
\end{algorithmic}
\end{algorithm}

\begin{algorithm}
\caption{Lexicographic Unranking Algorithm}\label{Unrank2}
\begin{algorithmic}[1]
\Procedure{LexUnrank}{$r,k,n$}\Comment{subset index, subset size, set size}
\State $x=1$
\For{$i=1:k$}
\While{$r \geq \binom{n-x}{k-i}$}
\State $r=r-\binom{n-x}{k-i}$
\State $x=x+1$
\EndWhile
\State $t_i=x$
\State $x=x+1$
\EndFor
\EndProcedure
\State \textbf{return} $T=(t_1, t_2, \ldots , t_k)$
\end{algorithmic}
\end{algorithm}

\begin{figure}
\hspace{-5mm}
    \begin{subfigure}[t]{0.4\textwidth}
    \centering
$$\begin{array}{r|cccc}
    d \setminus n& 2 & 3 & 4 & 5 \\
    \hline
    2 & \times & 4 & 6 & 12 \\
    3 & 4 & 9 \\
    4 & 5 \\
    5 & 6 \\
    6 & 8 \\
    7 & 9
\end{array}$$
\caption{Minimum Size}
    \end{subfigure}
    \begin{subfigure}[t]{0.4\textwidth}
    \centering
        $$\begin{array}{r|cccc}
        d \setminus n & 2 & 3 & 4 & 5  \\
        \hline
        2 & \times & 0.174 & 0.766 & 7842.597 \\
        3 & 0.231 & 108.606 \\
        4 & 0.664 &  \\
        5 & 2.342 \\
        6 & 493.003 \\
        7 & 36149.704
\end{array}$$
\caption{Runtime (sec)}
\end{subfigure}
\caption{Results for $\mathcal{B}(d,n)$ obtained using parallel computing and the corresponding runtime.}\label{HPCRes}
\end{figure}

\subsection{SMT Solvers}\label{Stan}

Satisfiability Modulo Theory (SMT) is a current area of research that is concerned with the satisfiability of formulas with respect to some background theory \cite{SMT}.  SMT solvers combine boolean SAT solving with decision procedures for specific theories.  For example, consider the following problem.  $$\begin{array}{ccc} a=b+1, & c<a, & c>b\end{array}$$ In the theory of the integers, this problem is not satisfiable (there are no integers a, b, c where all the expressions are true), however in the theory of the real numbers it is satisfiable (for example, with a=11, b=10, c=10.5).  In general, solving an SMT problem consists of first solving a SAT problem, then doing theory-specific reasoning, and then possibly going back and changing the SAT problem.  This process is repeated if necessary.  In addition, multiple theories can also be used in the same SMT problem instance, which may require additional repeats of this method.

To use SMT solvers on our identifying code problem for the undirected de Bruijn graph, we must first come up with a formulation of the problem using decision procedures.  The graph $\mathcal{B}(d,n)$, contains $d^n$ nodes.  For each of these, we create a boolean variable that denotes whether or not the node is part of the identifying code.  We then also create an array of boolean variables for that node's identifying set.  An assertion is added to make sure that each element of the array is true if and only if the corresponding neighbor's boolean variable is true (i.e. if and only if the neighbor is part of the identifying code).  To ensure unique codes, we add a statement to require that each node's identifying set is unique from every other node's identifying set.  Then, to get codes of a fixed size, we create an integer variable for each node and add the constraints that the integer is at least 0 and no greater than 1.  Next we add an assertion that each node's integer variable is 1 if and only if its boolean variable is true.  Finally, we add a constraint that the sum of all of the integer variables is equal to the desired identifying code size.

Now that the formulation of the problem has been determined, we can use a commercial SMT solver to find solutions.  For this work, we used the solver Z3, made by Microsoft Research.  We begin by first picking a code size, and asking if there exists an identifying code of that size.  If not, then the code size is increased by 1 and the problem is posed to Z3 again.  This continues until an identifying code of a specific size is found.  To find \textit{all} satisfying models, after a single model was found an assertion is inserted into the formulation that requires that the the previously found identifying code be eliminated as an option.  This forces Z3 to produce a different solution, or to state that the formulation is unsatisfiable (and hence no more identifying codes of that size exist). This process is repeated in a loop to obtain all identifying codes.

Using this approach on a single core, we were able to reproduce our results for $\mathcal{B}(d,n)$ from the HPC method in much less time.  See Figure \ref{SMTRes} for a summary of these results.  The numbers in parentheses denote that we found a code of that size, but did not eliminate the possibility of a smaller code existing.  The times given are determined by the time required to find one solution of minimum size in addition to the time required to determine that no smaller code exists.

\begin{figure}[t!]
\hspace{-5mm}
    \begin{subfigure}[t]{0.4\textwidth}
    \centering
    \small
    $$\begin{array}{r|cccccc}
        d \setminus n & 2 & 3 & 4 & 5 & 6 & 7 \\
        \hline
        2 & \times & 4 & 6 & 12 & (24) & (110)\\
        3 & 4 & 9 \\
        4 & 5 & 15 \\
        5 & 6 \\
        6 & 8 \\
        7 & 9 \\
        8 & (10)
\end{array}$$
\caption{Minimum Size}
    \end{subfigure}
    \hspace{20mm}
    \begin{subfigure}[t]{0.4\textwidth}
    \centering \small
        $$\scriptscriptstyle\begin{array}{r|cccc}
        d \setminus n & 2 & 3 & 4 & 5  \\
        \hline
        2 & \times & 0.118 & 0.490 & 20.800 \\
        3 & 0.038 & 5.415 \\
        4 & 0.178 & 37349.113  \\
        5 & 3.585 \\
        6 & 335.738 \\
        7 & 3707.735
\end{array}$$
\vspace{4mm}
\caption{Runtime (sec)}
\end{subfigure}
\caption{Minimum identifying codes on $\mathcal{B}(d,n)$ and the corresponding runtime for SMT solvers.}\label{SMTRes}
\end{figure}

Because of the advancements in current SAT and SMT solvers, they offer the potential to scale much better than a parallelized brute force approach.  This is due in part to the fact that many of today's solvers are capable of realizing which subsets of assignments will define an unsatisfiable result, and hence they will avoid models in which those statements are set.  In our problem, this might correspond to a case in which nodes $A$ and $B$ have the same identifying set.  In this case, the solver would not bother looking at combinations of True/False assignments on the other nodes that do not affect the identifying sets of $A$ or $B$.

In addition to the sophistication of today's solvers, there is also the possibility of parallelizing the search.  While some instances were run manually in a parallel manner for this experiment, there is some research to be done on automatically parallelizing the search in order to further our known minimum results.

\section{Conclusions}

The various methods discussed provide pure mathematicians with a range of opportunities for collaboration with scientists from several disciplines, such as computer science and physics.  The methods explored in this study included a parallel computing approach using Matlab, an adiabatic quantum optimization approach using a D-Wave quantum annealing processor, and lastly using satisfiability modulo theory (SMT) and corresponding SMT solvers.  From the base cases that we constructed using our variety of approaches, several new conjectures have been developed and eventually proven true \cite{BoutinHoran}, however the leg work needed to compute the base cases required a deep understanding of various computing techniques.

\section*{Acknowledgments}
Work using the D-Wave quantum annealing machine was performed jointly by AFRL/RI and Lockheed Martin under Air Force Cooperative Research and Development Agreement 14-RI-CRADA-02.

S. Adachi was supported by Internal Research and Development funding from Lockheed Martin.
S. Adachi would also like to thank Todd Belote and Dr. Andy Dunn of Lockheed Martin for their assistance respectively with the SAT-to-Ising mapping in Section \ref{subsec:SAT2Ising}, and with the generation of models for the scaling analysis of larger cases in Section \ref{subsec:Scaling}.

LOCKHEED MARTIN and LOCKHEED are registered trademarks in the U.S. Patent and Trademark Office owned by Lockheed Martin Corporation.

\bibliographystyle{amsplain}

\begin{thebibliography}{9}

\bibitem{Gadgets}
	R. Babbush, B. O'Gorman, and A. Aspuru-Guzik,
	``Resource Efficient Gadgets for Compiling Adiabatic Quantum Optimization Problems'',
	arXiv:1307.8041v1 [quant-ph]

\bibitem{SMT}
    C. Barrett, R. Sebastiani, S.A. Seshia, and C. Tinelli,
    ``Satisfiability Modulo Theories'',
    in \uline{Handbook of Satisfiability}, Chapter 26 (2009), 825-885.

\bibitem{Boixo}
	S. Boixo, T. Albash, F.M. Spedalieri, N. Chancellor, and D.A. Lidar,
	``Experimental signature of programmable quantum annealing'',
	\textit{Nature Communications} \textbf{4}:2067 (2013).

\bibitem{BoutinHoran}
    D. Boutin, V. Horan, and M. Pelto,
    ``Identifying Codes on Directed De Bruijn Graphs'', arXiv:1412.5842v2 (submitted).

\bibitem{Cai}
	J. Cai, W.G. Macready, A. Roy,
	``A practical heuristic for finding graph minors'',
	arXiv:1406.2741v1 [quant-ph]

\bibitem{NPHard}
    I. Charon, O. Hudry, and A. Lobstein,
    ``Minimizing the Size of an Identifying or Locating-Dominating Code in a Graph is NP-Hard'',
    \textit{Theoret. Comput. Sci.}, \textbf{290} (2003) no.~ 3, 2109-2120.

\bibitem{Choi1}
	V. Choi,
	``Minor-Embedding in Adiabatic Quantum Computation I. The Parameter Setting Problem'',
	\textit{Quantum Information Processing}, \textbf{7} (2008), 193-209.

\bibitem{Choi2}
	V. Choi,
	``Minor-Embedding in Adiabatic Quantum Computation II. Minor-universal graph design'',
	\textit{Quantum Information Processing}, \textbf{10} (2011), 343-353.

\bibitem{Farhi}
	E. Farhi, J. Goldstone, M. Sipser,
	``Quantum Computation by Adiabatic Evolution'',
	arXiv:quant-ph/0001106.

\bibitem{Gaitan}
	F. Gaitan and L. Clark,
	``Graph isomorphism and adiabatic quantum computing'',
	\textit{Phys. Rev. A}, \textbf{89}:2 (2014), 022342.

\bibitem{UndirIDCodes}
    V. Horan,
    ``On the Existence of $t$-Identifying Codes in Undirected De Bruijn Graphs'',
    arXiv:1508.00403.

\bibitem{Johnson}
	M.W. Johnson, M.H.S. Amin, S. Gildert, T. Lanting, F. Hamze, N. Dickson, R. Harris, A.J. Berkley, J. Johansson, P. Bunyk, E.M. Chapple, C. Enderud, J.P. Hilton, K. Karimi, E. Ladizinsky, N. Ladizinsky, T. Oh, I. Perminov, C. Rich, M.C. Thom, E. Tolkacheva, C.J.S. Truncik, S. Uchaikin, J. Wang, B. Wilson, and G. Rose,
	``Quantum annealing with manufactured spins'', \textit{Nature} \textbf{473} (2011), 194-198.

\bibitem{first}
    M.G. Karpovsky, K. Chakrabarty, and L.B. Levitin,
    ``On a New Class of Codes for Identifying Vertices Graphs'',
    \textit{IEEE Trans. Inf. Theory}, \textbf{355} (1998) no.~ 2, 599-611.

\bibitem{King}
	A.D. King, C.C. McGeoch,
	``Algorithm engineering for a quantum annealing platform'',
	arXiv:1410.2628v2 [cs.DS]

\bibitem{Klymko}
	C. Klymko, B. Sullivan, and T. Humble,
	``Adiabatic Quantum Programming: Minor Embedding With Hard Faults'',
	\textit{Quantum Information Processing}, \textbf{13} (2014), 709-729.

\bibitem{algos}
    Donald L. Kreher and Douglas R. Stinson,
    ``Combinatorial Algorithms:  Generation, Enumerations, and Search'',
    in \uline{Discrete Mathematics and Its Applications}, Book 7, CRC Press (1998):  Boca Raton.

\bibitem{isingPaper}
    A. Lucas,
    ``Ising Formulations of Many NP Problems'',
    \textit{Frontiers in Physics}, \textbf{2}:5 (12 Feb 2014).

\bibitem{Messiah}
	A. Messiah,
	\textit{Quantum Mechanics}, Vol. II, Wiley (1976): New York.

\bibitem{WSN}
    V. Mishra, J. Mathew, and D.K. Pradhan,
    ``Fault-Tolerant de Bruijn Graph Base Multipurpose Architecture and Routing Protocol for Wireless Sensor Networks'',
    \textit{Int. J. Sensor Networks}, \textbf{10}:3 (2011), 160-175.

\bibitem{chips}
    H. Moussa, A. Baghdadi, and M. Jezequel,
    ``Binary de Briujn onchip Network for a Flexible Multiprocessor LDPC Decoder'',
    \textit{ACM/IEEE Design Autmoation Conf.} (2008), 429-434.

\bibitem{Perdomo-Ortiz}
	A. Perdomo-Ortiz,J. Fluegemann, R. Biswas, and V.N. Smelyanskiy,
	``A Performance Estimator for Quantum Annealers: Gauge selection and Parameter Setting'',
	arXiv:1503.01083v1 [quant-ph]

\bibitem{Sensors}
    S. Ray, D. Starobnski, A. Trachtenberg, and R. Ungrangsi,
    ``Robust Location Detection with Sensor Networks'',
    \textit{IEEE J. Sel. Areas Commun.}, \textbf{22}:6 (2004), 1016-1025.

\bibitem{Rieffel}
	E.G. Rieffel, D. Venturelli, B. O'Gorman, M.B. Do, E. Prystay, and V.N. Smelyanskiy,
	``A case study in programming a quantum annealer for hard operational planning problems'',
	arXiv:1407.2887v1 [quant-ph]

\bibitem{ICDG}
    Y.C. Xu and R.B. Xiao,
    ``Identifying Code for Directed Graph,''
    in \uline{Software Engineering, Artifical Intelligence, Networking, and Parallel/Distributed Computing}, 2007.

\bibitem{Zick}
	K.M. Zick, O. Shehab, and M. French,
	``Experimental quantum annealing: case study involving the graph isomorphism problem'',
	arXiv:1503.06453v1 [quant-ph]

\end{thebibliography}

\end{document}